# Revisiting the Question of Information Content of EXAFS Spectra through a Bayesian Approach

Lucy Haddad*

*Queen Mary University of London and Diamond Light Source*

Diego Gianolio

*Diamond Light Source*

Andrei Sapelkin[†]

*Queen Mary University of London*

(Dated: September 9, 2025)

Over the last several decades the Shannon-Nyquist criterion has been widely used as a measure of the maximum information content in EXAFS spectra and provided an upper limit on the number of parameters used in fitting data. However, the criterion implicitly assumes independent parameters which is never the case in EXAFS analysis. Here we introduce a new criterion to measure the information content in EXAFS based on Bayesian approach that lifts the above condition. We test the new criterion by fitting the EXAFS spectrum of liquid gallium and demonstrate that not only it does constitute a superior measure of the data information content, but can also provide guidance in data analysis to differentiate between various fitting strategies.

## I. INTRODUCTION

X-ray absorption spectroscopy is a widely used technique for structural analysis of materials. Structural data (i.e. interatomic distances, coordination numbers, etc.) are typically extracted from the extended x-ray absorption fine structure (EXAFS) by fitting the model spectra to the experimental data using initial guess structural parameters that evolve during the fit to yield the required structural information. In practice, the process can be described as "to find the best model parameters that fit the data" using the following model function that describes an EXAFS signal [1]:

$$\chi(k) = \sum_j \frac{N_j S_0^2}{k R_j^2} F_j(k) e^{-2R_j/\lambda_j(k)} e^{-2k^2 \sigma_j^2}$$
$$\times \sin\left[2kR_j + \Phi_j(k)\right] \quad (1)$$

with

$$k = \frac{2\pi}{\lambda} = \sqrt{\frac{2m_e(E-E_0)}{\hbar^2}} \quad (2)$$

where $k$ is wavenumber, $\lambda$ is wavelength, $m_e$ is the electron mass, $E$ is the photoelectron energy, $E_0$ is the absorption edge position. The sum in Eq. 1equation.1.1 is taken over scattering paths $j$ originating from the absorbing (central) atom. The $N_j$ is the number of atoms in the scattering path, $R_j$ is the scattering path length, $\sigma_j^2$ describes mean-square displacements of atoms relative to the central atom, $F_j(k)$ is the photoelectron scattering amplitude, $\Phi_j(k)$ is the scattering phase, $\lambda_j(k)$ is the photoelectron inelastic mean free path, $S_0^2$ is an amplitude reduction factor accounting for relaxation of the absorbing atom due to the presence of the empty core level and multi-electron excitations. In a typical data analysis routine Eq. 1equation.1.1 is fitted to the experimental data by varying adjustable parameters: these are typically $N_j$, $R_j$, $E_0$, $S_0^2$ and $\sigma_j^2$ but may also include higher moments particularly at temperatures higher than the Debye temperature of the material of the otherwise harmonic (described by $\sigma_j^2$) interatomic displacements [2], [3], [4]. Other parameters (i. e. $F_j(k)$ and $\Phi_j(k)$) are usually calculated using a suitable ab initio code (e.g. FEFF [5], [6]) or obtained from tabulated values ($\lambda_j(k)$).

The relatively short range of EXAFS spectra significantly restricts EXAFS information content. The problem has been long recognised and received treatment in seminal work by Lee et. al. [7], based on information theory [8], giving the number of independent parameters that can be extracted from the data:

$$N_{ind} \approx \frac{2\Delta k \Delta R}{\pi} \quad (3)$$

where $\Delta k$ is the spectral range in momentum ($k$) space and $\Delta R$ is the corresponding range in $R$-space. This was further extended by Stern [9] to demonstrate that the exact expression is:

$$N_{ind} = \frac{2\Delta k \Delta R}{\pi} + 2 \quad (4)$$

and has become an accepted wisdom in EXAFS analysis, setting the upper limit on the number of parameters used in a fitting Eq. 1equation.1.1 to the data, and hence the information content of an EXAFS spectrum. The typical $\Delta k$ range in EXAFS is anywhere between 5 Å$^{-1}$ and 15 Å$^{-1}$ with the corresponding $\Delta R$ range between 1 Å and 5 Å, giving $N_{ind}$ between around 5 and 50 (and is generally closer to the lower estimate). However, Eqs 3equation.1.3 and 4equation.1.4 are obtained within two assumptions: i) Gaussian distribution of experimental errors; ii) parameters are independent. In practice, neither of these are true in the case of EXAFS analysis. Indeed, the first assumption implies that EXAFS signal is treated is "ground truth", ignoring any systematic errors potentially introduced by the procedure of the background subtraction to isolate the signal. The second is also an immediate issue in case of EXAFS where parameters are known to be correlated [10]. There are also experimental factors to consider that may introduce systematic errors, namely arising from detector non-linearities, sample non-uniformity, thickness and concentration variations [11], [12], [13].

---

* l.haddad@diamond.ac.uk
[†] a.sapelkin@qmul.ac.uk

All this calls into question to what extent Eq. 4equation.1.4 is appropriate for the description of the information that can be extracted from EXAFS data via multi-parameter fitting. The issue of parameter correlation has been long recognised [10] and suggestions were made [7] that eigenvalues and eigenvectors of the Hessian matrix of the fitting function must be analysed to assess and verify the validity of the fit and hence of the extracted information. However, in practice such analysis is almost never carried out and the question remains at what point one should stop introducing new parameters into a fit and to what extend the Eq. 4equation.1.4 (currently used to determine information content of EXAFS spectra) is appropriate in case of parameter correlation.

We have recently introduced a new figure of merit (FoM) in EXAFS analysis to characterise the goodness of fit and showed that it can be successfully used for structural model comparison [14]. Here we demonstrate that this new FoM, rooted in Bayesian statistical approach, can be used to address the problem of parameter correlation and to determine the number of parameters required to fit EXAFS data. With this, we are lifting the assumptions of Eq. 4equation.1.4 and provide the new measure of the information content in EXAFS.

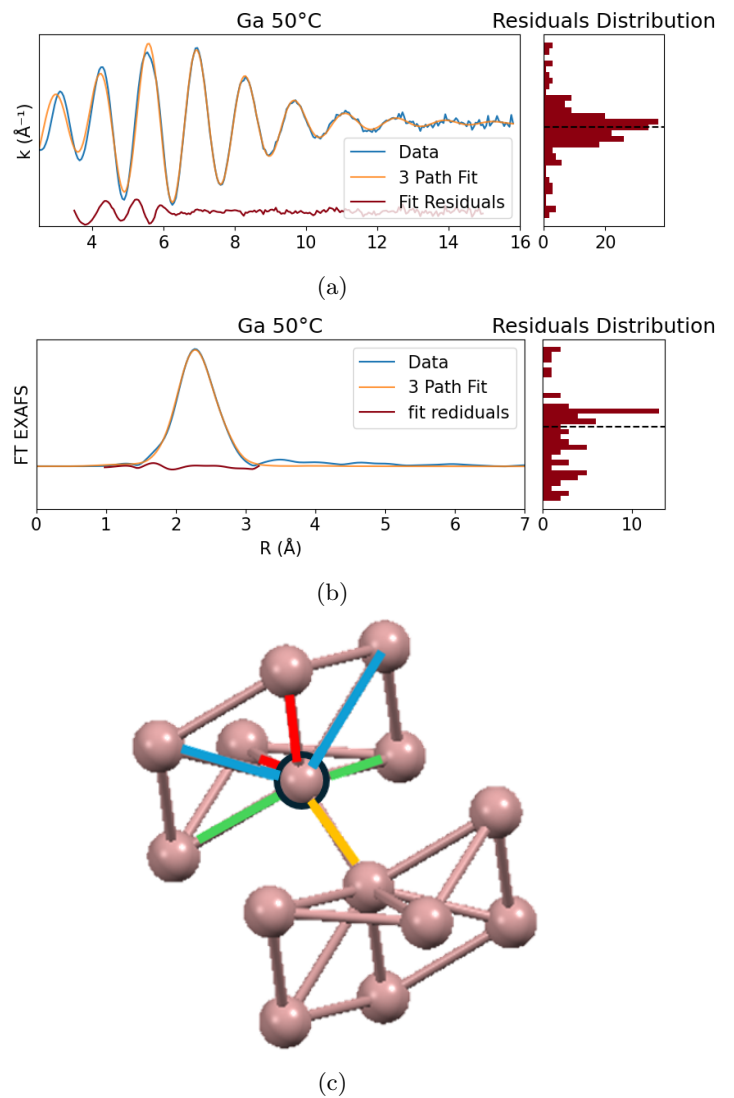

FIG. 1: Experimental data together with an example of fitting using Eq. 1equation.1.1, Table IModel names and their corresponding paths used to construct them.table.caption.7 using model 3b with default parameter ranges from "3 path fit", the dashed lines in (a) and (b) denote where residuals are equal to 0. (a) In k-space with k-weighting of 2. (b) In R-space. (c) The atomic structure of $\alpha$-Ga used to construct models.

## II. METHODOLOGY

As a test case to assess our analysis methodology we utilised EXAFS data for liquid gallium (l-Ga) at 50°C (melting temperature $T_{\rm m} = 29.8$°C). The typical EXAFS signal we used for analysis together with the corresponding magnitude of the Fourier transform (FT) can be seen in Fig. 1afigure.caption.5. We note that spectra with a single peak in FT are fairly typical in many EXAFS experiments, especially in experiments involving catalysis and nanoscale systems. Hence, we felt that the choice of a system with a well-studied structure while demonstrating a single prominent peak in FT would be most relevant (while can be easily extended to other systems). We modelled the local structure of l-Ga based on the crystalline $\alpha$-Ga [15, 16] shown in Figure 1cfigure.caption.5 with the following distances: single Ga-Ga pair at 2.48 Å (yellow), two Ga-Ga pairs at 2.69 Å (red), 2.73 Å (green) and 2.79 Å (blue) altogether giving 7 nearest neighbours.

In analysing the data we followed the steps fairly common in EXAFS data analysis. Using an appropriate structural model a user would typically fit $E_0$ (the absorption edge position), $S_0^2$, and the variables correspond-ing to each of the scattering paths $j$: $\Delta R_j$ (the difference between the pathlengths of the model and that found from fitting the EXAFS data), $\sigma_j^2$ and $N_j$. The user can include as many paths as required (with the total number of variable used in the fitting process $N_{\text{var}} \leq N_{\text{ind}}$) to obtain the best fit.

A number of figures of merit (FoMs) are used in EXAFS analysis to measure the quality of the fit: $\chi^2$, R-factor, BIC, AIC, [17] [18] (for details see Section 1 of Supporting Information). However, in many cases these FoMs do not give a clear answer as to whether to include new parameters into a fit or what is their significance. Hence, a heuristic user-adjustable "Happiness" parameter has been provided to address these issues [19]. It is based on decades of EXAFS analysis experience and includes, with varying (and user adjustable) weighting, penalties on the R-factor, parameter correlations, restraints and degrees of freedom in the fit. It is recognised as an important guide for data fitting but with no firm basis in statistics it cannot be quoted in publications.

We have recently introduced [14] a new FoM for comparison of structural models in EXAFS — Bayes Factor Integral (BFI). This FoM is based on Bayesian approach to statistical data analysis and we demonstrated that it can be used in the analysis of nanoparticles to select the best structural model that fits the data in EXAFS. This new FoM is defined as follows [14]:

$$BFI = (2\pi)^{m/2} L_{\max} \frac{\sqrt{\det(\mathbf{Cov_p})}}{\prod_{i=1}^{m} \Delta p_i} \quad (5)$$

where $m$ is the number of parameters in the model, $\Delta p_i$ are the initial fitting parameter ranges, $\mathbf{Cov_p}$ is the parameter covariance matrix, and $L_{\max}$ is the likelihood for the fit. This form of the $BFI$ is similar to the expression for the global likelihood for the model parameters [20], except that $\mathbf{Cov_p}$ need not be diagonal [14]. Thus, although falling short of the full analysis of the Hessian matrix, it does take into account parameter correlations, while returning a numerical estimate for the best fit. Furthermore, this new FoM automatically quantifies the so-called *Occam's razor* by including the prior model information (in the form of parameter ranges $\Delta p_i$) in the denominator and together with the $\mathbf{Cov_p}$ gives the *Occam factor*:

$$\Omega = (2\pi)^{m/2} \frac{\sqrt{\det(\mathbf{Cov_p})}}{\prod_{i=1}^{m} \Delta p_i}. \quad (6)$$

Occam's razor is a problem-solving principle that suggests searching for a model constructed with fewest possible parameter set that satisfies the data. In the case of EXAFS thus formulated FoM addresses the problem of parameter correlation through the covariance matrix and allows for prior information about the structure (in the form of parameter ranges) to be included, thus benefiting the fits with little correlation between parameters and better prior information about the model. In practice, the values of $BFI$ can vary drastically, hence we use $\ln BFI$ to quantify the goodness of fit [21], [14]. When using the BFI as a FoM the $\ln BFI$ values are compared between the models (e.g. model 1 and model 2) by defining the Bayes Factor (BF) $\ln BF = \ln BFI_1 / \ln BFI_2 = \ln(BFI_1 - BFI_2)$ with the following a scale for how to pick one model over another [14], [22]:

- $< 1$ – barely worth considering,
- $1 - 2$ – substantial,
- $2 - 5$ – strong evidence,
- $> 5$ – decisive.

All data was fit in Larch [17]. For the calculation of $\chi^2$, $\chi_\nu^2$ and subsequently parameter errors in Larch either $R$- or $k$-space noise is used to re-scale these values [23]. In Larch the k-space noise estimate ($\epsilon_k$) is often an underestimate [24], it is indirectly calculated from $R$-space noise (See Appendix Asection*.17). For our reported FoMs and parameter errors we used the standard deviation of the EXAFS signal between $14.0 - 15.0$ Å$^{-1}$ which gave $\epsilon_k \approx 0.00996$ Å$^{-1}$. In all cases we fit the signal corresponding to the first peak in $R$-space (Figure 1bfigure.caption.5) at around 2.5 Å. The model structure (Figure 1cfigure.caption.5) has 3 groups of two atoms at the same respective distances from the central Ga atom and one at a shorter distance as described above. This has been split into 4 separate single-scattering paths contributing to the single peak shown in FT (Fig. 1bfigure.caption.5).

We started by fitting the data with a single path corresponding to a model in which all nearest-neighbour distances are equal. This is a logical first step, considering only a single peak is observed in the EXAFS FT (Fig. 1bfigure.caption.5). Here we used 4 single path models with the distances between $2.48 - 2.79$ Å (see Models 1a–d in Table IModel names and their corresponding paths used to construct them.table.caption.7); these are models with 4 free parameters: $E_0, \Delta R, \sigma^2, N$ (with $S_0^2$ being fixed at $S_0^2 = 1$ [25]). Once all different single-path models have been fit and FoMs calculated, the data is fitted with a 2-path model with 7 free parameters. This corresponds to 2 sets of nearest-neighbour distances from the absorbing Ga atom (models 2a–2f in Table IModel names and their corresponding paths used to construct them.table.caption.7). Once all possible 2-path permutations have been fit and the $BFI$ and standard FoMs calculated, then a $3^{rd}$ path is added in. These are models 3a–3c and have 10 free parameters. In cases where 10 parameters is greater than $N_{\text{ind}}$ in Eq. 4equation.1.4, coordination number for one of the paths is fixed (see models 3d and 3e in Table IModel names and their corresponding paths used to construct them.table.caption.7). To test whether the $BFI$ would equally favour two different models with the same number of parameters, models 2g and 2h were used. These have 9 free parameters (the same as models 3d and 3e) but use $3^{rd}$ cumulants added instead of a $3^{rd}$ scattering path. Initially the data were fit in the $3.50 \leq k \leq 15.00$ Å$^{-1}$ k-range, which gives $N_{\text{ind}} = 11$ (according to Eq. 4equation.1.4). After this the k-range was systematically reduced to $3.50 \leq k \leq 13.00$ ($N_{ind} = 10$) and then it was reduced to $3.50 \leq k \leq 12.00$ ($N_{ind} = 9$) in order to evaluate the sensitivity of $BFI$ to the reduced data range.

| Paths Used | Name | Number of Parameters |
|---|---|---|
| 269 | 1a | 4 |
| 273 | 1b | 4 |
| 279 | 1c | 4 |
| 248 | 1d | 4 |
| 269 + 273 | 2a | 7 |
| 269 + 279 | 2b | 7 |
| 269 + 248 | 2c | 7 |
| 273 + 279 | 2d | 7 |
| 273 + 248 | 2e | 7 |
| 279 + 248 | 2f | 7 |
| $273 + 248 + C3_{273}$ | 2g | 8 |
| $273 + 248 + C3_{248}$ | 2h | 8 |
| $273 + 248 + C3_{273} + C3_{248}$ | 2i | 9 |
| 273 + 248 + 279 | 3a | 10 |
| 273 + 248 + 269 | 3b | 10 |
| 273 + 269 + 279 | 3c | 10 |
| $273 + 248 + 269|_N$ | 3d | 9 |
| $273 + 248 + 279|_N$ | 3e | 9 |
| $273 + 248 + 279 + C3_{273}$ | 4a | 11 |

TABLE I: Model names and their corresponding paths used to construct them.

Using this approach we investigated the dependence of $\ln(BFI)$ as a function of number of parameters, compared the result with that obtained from Eq. 4equation.1.4 and demonstrate that $BFI$ can be used to provide the limit on the information content of the data together with the methodology to select between various data fitting strategies and structural models, thus constituting a universal FoM in EXAFS data analysis.

## III. RESULTS AND DISCUSSION

Following the data fitting methodology outlined in the previous section for the structural model of l-Ga, we compiled a table of scattering paths with an increasing number of variables - these are shown in Table IModel names and their corresponding paths used to construct them.table.caption.7. For the ranges ($\Delta p_i$ terms in Eq. 5equation.2.5) we set them close to typical fitting bounds on parameters found in Larch: $E_0 \pm 10$ eV, $\Delta R$ was set to $\pm 0.25$Å, $\Delta N$ was set to $\pm 3.5$ which is 50% of total Ga absorbing atom coordination number and $\Delta \sigma^2 \pm 0.02$ Å$^2$.

$E_0$ is a shift in the edge position accounting for errors in experimental calibration and for empirical convention in determination of the absorption edge position and its range typically does not exceed 10 eV so its corresponding range for the BFI calculation was set to be this, [26]. The $\Delta R$ range initially was $\pm 0.25$ and was later reduced to $\pm 0.025$. The initial range means that interatomic distances can vary up to 25% of their initial value. This is half of the default $\Delta R$ fitting bound in Larch and was set at this large value since we measure the EXAFS spectrum of l-Ga so distances in its structure could be significantly different from c-Ga which our model was derived from. This was later found to be too large of a range because allowing bond distances to vary by this much actually could allow paths to duplicate or switch within this range. So it was then reduced to give a more meaningful comparison of models. Atomic displacement $\sigma^2$ can be calculated using Debye or Einstein approximations and was calculated for c-Ga to be around 0.02 Å$^2$ at room temperature.[2, 27, 28].

The total number of central Ga nearest-neighbours in c-Ga is 7, so the range on $N$ was set to be half of this meaning fitted coordination numbers can vary within half of this number.

### A. Method with Default Background Subtraction

For the first three results with the default $\Delta p_i$ terms used, the $\ln(BFI)$ are shown in Figures 2afigure.caption.9–2dfigure.caption.9. When the data range is between $3.50 \leq k \leq 15.00$ Å$^{-1}$ the maximum number of parameters defined by Eq. 4equation.1.4 is 11. However, we found that the $\ln(BFI)$ has a maximum at 10 parameters, corresponding to model 3c which describes the first-shell FT signal as having 3 single scattering paths. When 11 parameter are used the $\ln(BFI)$ was found to drop significantly ($\ln(BF) \approx 10$) indicating that 10 parameter fit is decisively better.

When the fitting range is reduced to $3.50 \leq k \leq 13.00$ Å$^{-1}$ such that maximum number of parameters is 10 (Figure 2bfigure.caption.9) the $\ln(BFI)$ reaches a maximum at this limit. The two models ranking the highest are now 3c and 3b with $\ln(BF) < 1$ between the two models, meaning neither is favoured significantly over the other. Still, these the 3-paths models are favoured significantly ($\ln(BF) \approx 7$) over 1- or 2-path ones here also, so the effect of the reduced data range is to make the 3-paths models indistinguishable.

When the data range is reduced again to between $3.50 \leq k \leq 12.00$ Å$^{-1}$ giving $N_{ind} = 9$ we can no longer use the same fitting approach for 3-paths models as the number of variables would be larger than $N_{ind}$ (see Table IModel names and their corresponding paths used to construct them.table.caption.7). Here we used two methods to address that. The first (see Figure 2cfigure.caption.9) involved fitting of up to 2 paths, followed by introduction of 3rd path in model 3d (and 3e) with the coordination number fixed for that path. This was in order to fit up to three paths but still remain within the parameter limit since in Larch covariance is no longer calculated once $N_{var} > N_{ind}$ (and hence $BFI$ cannot be calculated). We found that a 3-path model again is the highest ranked, with model 3d having a substantial difference in $\ln(BFI)$ over 3e.

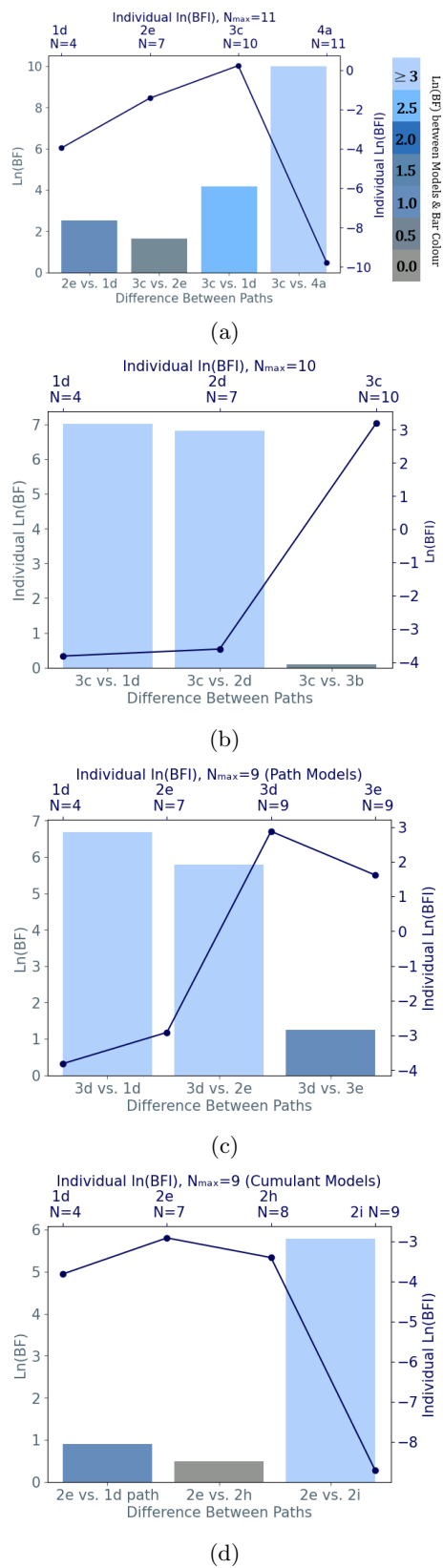

FIG. 2: $\ln(BFI)$ versus number of fitted parameters for the Ga $50^oC$ spectrum with default parameter ranges. (a) In the full data-range with $N_{ind} = 11$, with the legend indicating choice of colour for the bars with respect to the $ln(BF)$ between models. (b) In the first fitting-range reduction with $N_{ind} = 10$. (c) For the second fitting-range reduction with $N_{ind} = 9$: 3 path model. (d) With $N_{ind} = 9$: cumulant model.

Figure 2dfigure.caption.9 shows the analysis of the results when two alternative methods to reach the parameter limit are used. The number of parameters can also be increased by using higher moments of the distance distribution, rather than by addition of scattering paths. Here, 2 scattering paths were used and $3^{rd}$ cumulants were included and fitted into different paths to reach the limit. The ranges set on $3^{rd}$ cumulants were $\pm 0.1$ Å$^3$. In this case the number of parameters is peaking at $N = 7$ with models 1d, 2e and 2h performing similarly ($\ln(BF) < 1$) were favoured with no significance between either one. However, in comparison with the 3-path model in Figure 2cfigure.caption.9 the 2-path models $BFI$s are still substantially lower ($\ln(BF) > 5$) than the 3-path ones with coordination numbers fixed, indicating that the latter are the best fits to the data.

Thus, we see that at the full fitting range the $\ln BFI$ points to a 3-path model with fewer parameters than the maximum from Eq. 4equation.1.4. When the fitting range is reduced (to decrease the information content of the spectrum) the BFI consistently favours 3-path models regardless of how close to the information content limit they fall. In its crystalline state one would expect the first shell Ga EXAFS to be described by 4 single scattering paths- due to the 4 distances from the 7 atoms in the first shell of its crystalline structure [15]. For l-Ga it is known that average bond distance shorten, [16, 29]. So within our tolerance range set on the $\Delta R$ terms being fit ($\pm 0.25$Å) this could result in an effect that 3 single scattering paths adequately fit our l-Ga signal with the longest single scattering path no longer being relevant. We see this at all 3 fitting ranges where 3-path models are significantly favoured by the BFI than others. However here the specific combination of scattering paths constituting the 3-path model is not the same within all examined ranges with Models 3c, 3b and 3d are the highest ranked ones respectively for each data-range. The ranges used for $\Delta p_i$ and the fitting bounds in Larch were left large - $\pm 0.25$Å. Within this range paths could duplicate or interchange giving no significance to which paths are listed under the model. Hence, the same methodology for the full spectrum was repeated with the reduced $\Delta R$ ranges of $\pm 0.025$ Å.

With the more constrained fitting bounds on $\Delta R$ the same method of fitting paths and cumulants up to the parameter limit was repeated. The results of $\ln(BFI)$ are shown in Figure 3$\ln(BFI)$ versus the number of parameters for the full data range, $N_{ind} = 11$ with tighter constraints on the fitting parameters.figure.caption.10. The BFI has a maximum at $N = 10$, again, below the $N_{ind}$. This maximum corresponds to model 3b, a 3-path model describing the first peak in EXAFS FT. There is a large drop in model significance once a $3^{rd}$ cumulant is added to the fit, these results are consistent with the first fit in Figure 2afigure.caption.9 but with larger $\ln(BFI)$ differences between models. This is because the models are the same but fitting bounds are constrained meaning paths that do not fit the data will be more heavily penalized by the BFI.

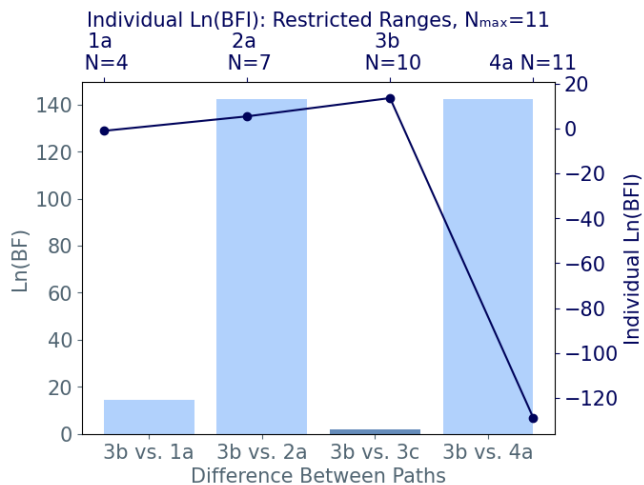

FIG. 3: $\ln(BFI)$ versus the number of parameters for the full data range, $N_{ind} = 11$ with tighter constraints on the fitting parameters.

For the full data-range with both the initial and restricted ranges a 4-path model was also tested. The model involves single scattering corresponding to the 4 paths (at $2.48, 2.69, 2.73$ and $2.79$ Å, designated by colours in Fig. 1cfigure.caption.5) in the first coordination shell around the central atom. In this case the of the scattering paths with the lowest amplitudes were fixed to keep the number of parameters in line with the maximum allowed by the Shannon-Nyquist criterion. The full results are given in Tables $VI$ and $XIX$ in the supplementary material. The results for the 4-path model follow the previously demonstrated trend in Figures 2afigure.caption.9 and 3ln($BFI$) versus the number of parameters for the full data range, $N_{ind} = 11$ with tighter constraints on the fitting parameters.figure.caption.10 with the $BFI$ values for the 4-path models being significantly lower than for 3-path while being higher than the cumulant models.

Thus, we find that BFI provides an clear metric for the data information content and model selection. In the case of l-Ga we can see that the structural model used in the fit must include at least 3 distinct scattering paths, while higher order cumulants are not favoured. This may appear odd, considering the contribution of higher order cumulants tends to play a more prominent role at elevated temperatures due to increase in anharmonicity of atomic motions [4]. However, l-Ga is known to exhibit a number of interesting anomalies including reduction in NN distance [30] and increase in covalency as a function of temperature [30, 31]. Each of these would result in an increase in the correlations of the local atomic motions between NN, potentially reducing the anharmonic effects. Hence, our analysis may also provide important insights into the physics of the system behaviour on atomic scale.

Having established the utility of BFI as a measure of the data information content in a typical data analysis procedure, we next tested the impact of the background subtraction on the BFI-based model selection.

### B. The Case of Poor Background Subtraction

As the next test of BFI, the fitting methodology from Section III Asection*.8 was applied to the l-Ga data with poor background subtraction. The $3.50 \leq k \leq 15.00$ Å$^{-1}$ data range was used ($N_{ind} = 11$) and both the default and restricted ranges on parameter bounds were tested. The data used together with an example fit are shown in Figure 4The $k^2$ (a) and FT EXAFS (b) for Ga at $50^oC$ with purposefully bad background subtraction (model 3b with default parameter ranges as the 3 path fit in Figure 1afigure.caption.5). Again the dashed lines denote where the residuals are equal to 0.figure.caption.12 where one can see significant departure from the histogram of the residuals seen in Figure 1afigure.caption.5 with the difference particularly clear to see in Figure 4The $k^2$ (a) and FT EXAFS (b) for Ga at $50^oC$ with purposefully bad background subtraction (model 3b with default parameter ranges as the 3 path fit in Figure 1afigure.caption.5). Again the dashed lines denote where the residuals are equal to 0.figure.caption.12b with the additional peak at around 1 Å.

The impact of bad background subtraction is clear from Figure 5afigure.caption.13 where the results of the analysis are shown for the default parameter ranges. We can see that the trend is now different from the results in Figs . 2afigure.caption.9 and 3ln($BFI$) versus the number of parameters for the full data range, $N_{ind} = 11$ with tighter constraints on the fitting parameters.figure.caption.10: the $\ln(BFI)$ returns maximum value at $N_{ind}$ with model 4a (a 3-path model with a 3rd cumulants) favoured over all others. This clearly shows that model selection is very sensitive to the quality of the data. Interestingly, we found that when the fitting bounds on $\Delta R$ are reduced to $\pm 0.025$ Å the trend from the previous section re-emerges. This is shown in Figure 5bfigure.caption.13 where we can see that again a 3-path model (3b) is significantly better (according to the BFI) than all others while the introduction of higher order cumulants causes a large drop in $ln(BFI)$. Looking at the structure of BFI in Eq. 5equation.2.5, it appears that reducing the parameter bounds also results in drastic reduction in parameter correlations. Thus, the overall message from this part is the crucial role of adequate background subtraction and importance of prior information (that is used to set parameter ranges) in the data analysis.

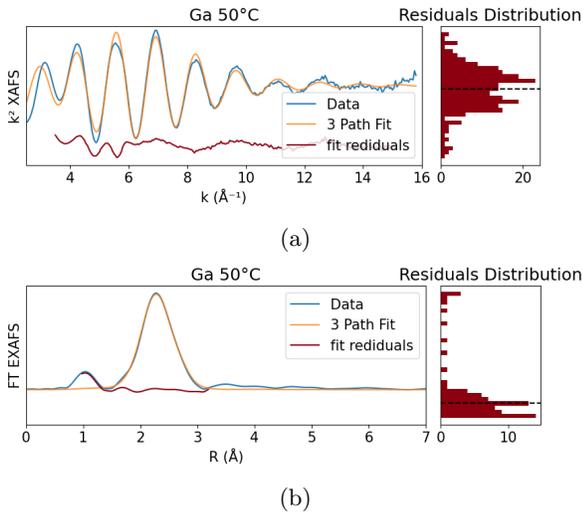

FIG. 4: The $k^2$ (a) and FT EXAFS (b) for Ga at $50^oC$ with purposefully bad background subtraction (model 3b with default parameter ranges as the 3 path fit in Figure 1afigure.caption.5). Again the dashed lines denote where the residuals are equal to 0.

## IV. CONCLUSIONS

The information content of (the number of variables that can be used to fit) EXAFS data has been defined by the Shannon-Nyquist criterion for several decades. However, the latter is only strictly applicable to the analysis of EXAFS data on a number of assumptions that are typically not met. Besides, the existing information criteria do not take into account prior information about the data (e.g. ranges of the fitting variables). To address that, here we introduced and tested a new criterion for the information content of EXAFS data based on BFI. This new approach lifts the assumption of independence of the fitting parameters used to obtain structural information from an EXAFS spectrum and allows to account for prior information about the models being tested. We show that the BFI-determined information content is different from that defined by the widely used Shannon-Nyquist criterion (that is strictly valid for independent parameters only). We tested the new approach for evaluating the maximum available information by fitting the EXAFS spectrum of l-Ga and found that BFI-based measure of information shows sensitivity to the variation in the data range and to the data quality. We demonstrated that our approach not only delivers a new measure of the data information content, but can also provide guidance in data analysis to differentiate between various fitting strategies, as well as yield insights into the physics of materials on atomic scale.

This work demonstrates the BFI approach provides for a versatile metric in data analysis, it informs the user when to stop adding parameters to the model and will point to the best model that fits the data. Furthermore, since BFI calculation are quite straightforward, this approach should lend itself easily to further automation, allowing testing multiple models. Another important point is that, being versatile, this new criterion can also be applied across a much wider range of spectra/signal analysis problems (e.g. Raman, IR, photolumenescence etc.) in the same systematic way demonstrated in this work.

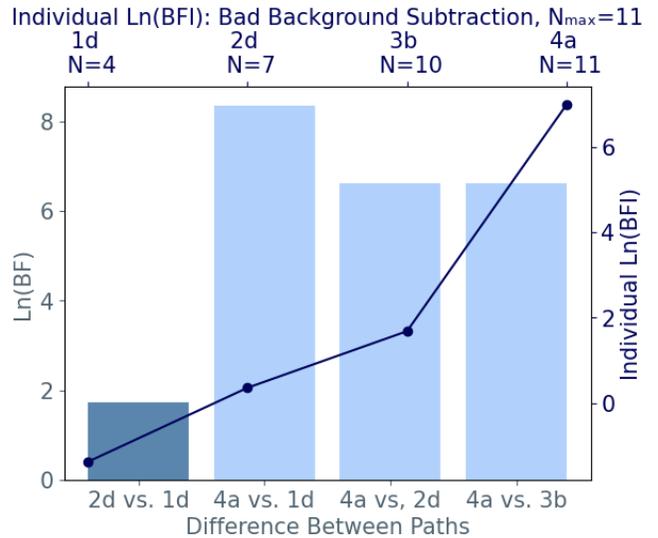

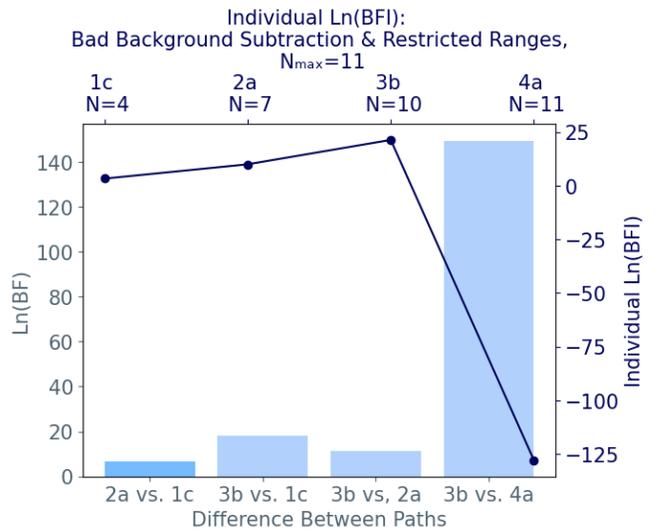

FIG. 5: $\ln(BFI)$ versus number of fitted parameters for the Ga $50^oC$ with poor background subtraction. (a) For the full data range. (b) For the full data range and restricted parameter ranges.

Full tables of results can be viewed in the supplementary material.

## AUTHOR CONTRIBUTIONS

Data acquisition was performed by D.G. and L.H.. A.S and L.H. performed the data analysis.

## ACKNOWLEDGMENTS

L.H. is grateful to Diamond Light Source and Queen Mary University of London for the joint studentship and funding to support this work.

## Appendix A: Figures of Merit in EXAFS

In IFFEFIT [18], the chi-squared ($\chi^2$) and reduced chi-squared ($\chi^2_\nu$) fitting metrics are calculated via Eqs. A1equation.A.1,A2equation.A.2:

$$\chi^2 = \frac{N_{ind}}{\epsilon N_{data}} \sum_{i=min}^{max} \left[ Re(\chi_d(r_i) - \chi_t(r_i))^2 + (\chi_d(r_i) - \chi_t(r_i))^2 \right] \tag{A1}$$

$$\chi^2_\nu = \frac{\chi^2}{\nu} \tag{A2}$$

$\nu$ is degrees of freedom in the EXAFS fit: $N_{ind} - N_{var}$. $\chi^2$ and $\chi^2_\nu$ are the most commonly reported figures of merit (FoMs) in EXAFS fitting and are the first on the list of FoMs in EXAFS fitting reports from programs such as Larch and Artemis [17][18]. They are a scaled measure of how closely the model ($\chi_t(r)$) and data ($\chi_d(r)$) fit.

$\epsilon$ is the measurement uncertainty and the R-space measurement of this is the average $\chi_d$ fluctuations in high-R regions ($\geq 15\text{Å}$) of the spectrum. This is because there is typically no significant XAFS signal past these points so the signal will be dominated by random fluctuations[23].

To obtain the uncertainty in k-space (we will call $\epsilon_k$) from R-space uncertainty ($\epsilon_R$) we assume the random fluctuations in both k- and R-space are "white noise" with constant root-mean-square amplitude and is constant over all values of $k, R$. $\epsilon_k$ and $\epsilon_R$ are related by Parseval's theorem [32]. For the symmetric XAFS transform of k-weighted spectrum with finite k-window between $k_{max}$ and $k_{min}$ we get from Parseval's theorem:

$$\int_{k_{min}}^{k_{max}} |\epsilon_k k^\omega|^2 dk = 2 \int_0^{\pi/2\delta k} |\epsilon_R|^2 dR \tag{A3}$$

Taking $\epsilon_R$ and $\epsilon_k$ to be constants we get:

$$\epsilon_k = \epsilon_R \sqrt{\frac{\pi(2\omega+1)}{\delta k(k_{max}^{2\omega+1} - k_{min}^{2\omega+1})}} \tag{A4}$$

These $\epsilon_k$ and $\epsilon_R$'s have no k- or R-dependence so they cannot account for systematic errors such as poor background removal, approximations and errors in theory, sample inhomogeneity and detector non-linearity [33]. So these values are a lower bound on the measurement uncertainty. But because of the simplicity of calculating these values they are easily automated and require no prior information regarding data quality.

The implications of $\epsilon$ being a lower estimate of actual uncertainty affect $\chi^2$ and $\chi^2_\nu$ values calculated from the fit. $\chi^2$ will be large and $\chi^2_\nu$ cannot be interpreted for a single fit.

The AIC and BIC have similar forms, both including the number of fitted parameters and maximum likelihood. $N$ refers to the number of estimated parameters, $n$ being the number of data-points being fit and $L_{max}$ being the maximum likelihood of the model.

$$AIC = 2N - 2Ln(L_{max}) \tag{A5}$$

$$BIC = NLn(n) - 2Ln(L_{max}) \tag{A6}$$

## Appendix B: BFI

$$BFI = (2\pi)^{m/2} L_{max} \frac{\sqrt{\det(\mathbf{Cov_p})}}{\prod_{i=1}^{m} \Delta p_i} \tag{B1}$$

From Bayes theorem: for a Model M which can be described as assigning/fitting a series of amplitudes $\{A_\alpha\}$ to a set of model basis functions $g$ (in which the complete set of $\sum g_\alpha A_\alpha$ will describe the model) to set of data points D we have (i.e. multi-parameter fitting):

$$P(\{A_\alpha\}|D, M) = \frac{P(D|\{A_\alpha\}, M) P(\{A_\alpha\}|M)}{P(D|M)} \tag{B2}$$

For model comparison we want to compute the odds ratio: probability of one model versus the other, a ratio of each models' probabilities:

$$O_{ij} = \frac{P(M_i|D,I)}{P(M_j|D,I)} = \frac{P(M_i|I)}{P(M_j|I)} \frac{P(D|M_i,I)}{P(D|M_j,I)} = \frac{P(M_i|I)}{P(M_j)|I)} BF_{ij} \tag{B3}$$

Where prior information is denoted I. $BF_{ij}$ is the Bayes Factor between the two models, a ratio of the two models' global (also called marginal) likelihoods.

$$BF_{ij} = \frac{P(D|M_i, I)}{P(D|M_j, I)} \tag{B4}$$

Our Bayes Factor is a ratio of the two models $M_i, M_j$'s marginal likelihoods. To get $P(D|M, I)$ we integrate over all possible values from the model, a **marginal likelihood integral**, or MLI. For this reason to avoid confusion with this calculation versus our FoM, we call our FoM the BFI, Bayes Factor Integral.

It is because of this Bayes factor that Bayesian statistics are often not used in general for model comparison. Calculation of a models' global (marginal) likelihood will involve integrating (marginalising) over all of the models' parameters from the product of the prior and the likelihood. It is shown that for the linear-model case[20] where errors are independent and normally distributed we get $P(D|A_\alpha, I)$ as a product of 2-D Gaussians:

$$P(D|\{A_\alpha\}, I) = \frac{1}{\sigma^N (2\pi)^{N/2}} \exp\left[\frac{-1}{2\sigma^2} \sum_{i=1}^N (d_i - f_i)^2\right]$$
$$= \frac{1}{\sigma^N (2\pi)^{N/2}} e^{\frac{-Q}{2\sigma^2}} \quad \text{(B5)}$$

Where Q is a sum of squared residuals:

$$Q(\{A_\alpha\}) = \sum_{i=1}^N (d_i - f_i)^2 \quad \text{(B6)}$$

$d_i, f_i$ are the values of the data and model at point i respectively.

We use a uniform prior such that:

$$P(\{A_\alpha\}|D, I) = \frac{1}{\prod_\alpha \Delta A_\alpha} \quad \text{(B7)}$$

where the denominator is a product of the ranges set on parameter values (we call them $\Delta p_i$ in the bfi equation). For global likelihood we now get:

$$P(\{A_\alpha\}|D, I) = \frac{P(\{A_\alpha\}|I)}{P(D|I)} e^{\frac{-Q^2}{2\sigma^2}}$$
$$= \int_{\Delta A} d^M A_\alpha P(\{A_\alpha\}|I) P(D|A_\alpha, I)$$
$$= \frac{1}{\prod_\alpha \Delta A_\alpha} \frac{1}{\sigma^N (2\pi)^{(N/2)}} \int_{\Delta A} d^M A_\alpha^{\frac{-Q}{2\sigma^2}} \quad \text{(B8)}$$

Where N is the number of parameters in our model and M is the number of data-points we are fitting. We can call $Q_{min}$ the parameter values at which residuals are minimized and write:

$$P(D|M_i, I) = \frac{1}{\prod_\alpha \Delta A_\alpha} \frac{1}{\sigma^N (2\pi)^{N/2}} e^{\frac{-Q_{min}}{2\sigma^2}} \int_{\Delta A} d^M A_\alpha e^{\frac{\Delta Q}{2\sigma^2}} \quad \text{(B9)}$$

Defining $\chi^2$ as $Q/\sigma^2$ and transforming to an orthonormal set of model basis functions we get the final results:

$$P(D|M_i, I) = \left[\frac{(2\pi)^{M/2} \sqrt{det((\mathbf{V}))}}{\prod_\alpha \Delta A_\alpha}\right] \frac{1}{\sigma^N (2\pi)^{N/2}} e^{\frac{-\chi^2_{min}}{2}} \quad \text{(B10)}$$

$\mathbf{V}$ is the parameter covariance matrix.

This can be written as the "Occam factor" multiplied by the model's maximum likelihood:

$$\Omega_M \hat{\mathcal{L}}_{max} \quad \text{(B11)}$$

$\hat{\mathcal{L}}_{max}$ is likelihood for model at the mode:

$$\hat{\mathcal{L}}_{max} = P(D|\hat{A}, M_i) = \frac{1}{\sigma^N (2\pi)^{N/2}} e^{\frac{-\chi^2_{min}}{2}} \quad \text{(B12)}$$

And Occam factor for the model is:

$$\Omega_M = \frac{(2\pi)^{M/2} \sqrt{det(\mathbf{V})}}{\prod_\alpha \Delta A_\alpha} \quad \text{(B13)}$$

In the non-linear case one can get the result using the Laplace approximation for the Bayes factor[34] that the global likelihood of a model is:

$$P(D|M_i, I) \approx P(\hat{\theta}|M_i, I) \hat{\mathcal{L}}_{max} (2\pi)^{M/2} det(\mathbf{I})^{-1/2} \quad \text{(B14)}$$

Where $\mathbf{I}$ is the information matrix and $det(\mathbf{I})^{-1} = det(\mathbf{V})$ so we get:

$$P(D|M_i, I) = \frac{(2\pi)^{M/2} \sqrt{det(\mathbf{V})}}{\prod_\alpha \Delta A_\alpha} \hat{\mathcal{L}}_{max} = \Omega_m \hat{\mathcal{L}}_{max} \quad \text{(B15)}$$